\documentclass[aps,preprint]{revtex4}

\usepackage{amsmath,graphicx}

\begin{document}

\title{
Replica trick with real replicas: A way to build in thermodynamic
homogeneity}


\author{V\'aclav Jani\v s} \author{Lenka Zdeborov\'a}%

\affiliation{
Institute of Physics, Academy of Sciences of the Czech Republic,\\ Na
Slovance 2, CZ-18221 Praha, Czech Republic}



\begin{abstract}{
We use real replicas to investigate stability of thermodynamic
homogeneity of the free energy of the Sherrington-Kirkpatrick (SK) model
of spin glasses. Within the replica trick with the replica symmetric ansatz
we show that the averaged free energy at low temperatures is not
thermodynamically homogeneous. The demand of minimization of the
inhomogeneity of thermodynamic potentials leads in a natural way to the
hierarchical solution of the Parisi type. Conditions for the global
thermodynamic homogeneity are derived and evaluated for the SK and
$p$-spin infinite range models.}
\end{abstract}

\maketitle

\section{Introduction}

Thermodynamic potentials have to be state variables.  It is so if they are
thermodynamically homogeneous. Thermodynamic homogeneity means that the
densities of thermodynamic potentials depend on the extensive variables
only via their densities. This property is usually expressed as the Euler
condition $\nu F(T,V,N,\ldots,X_i,\ldots) = F(T, \nu V, \nu N,\ldots,\nu
X_i,\ldots)$, where $\nu$ is an arbitrary positive number and $X_i$
exhaust all extensive variables. It is straightforward to obtain from the
Euler condition $f\equiv F/V=f(T,1,N/V,\ldots,X_i/V,\ldots)$. This linear
homogeneity condition leads to the existence of the thermodynamic limit
independently of boundary conditions and the shape of the volume.

One of the authors recently argued that simple solutions (replica
symmetric) of the SK model do not obey thermodynamic homogeneity
\cite{Janis03b}. Using the TAP approach a hierarchical solution equivalent
to the Parisi replica-symmetry breaking (RSB) was derived from the demand
of homogeneity, or better from minimizing the inhomogeneity of the
averaged free energy.

In this paper we show that the same demand of thermodynamic homogeneity can
be implemented into the replica trick for the SK and other mean-field
models as scale invariance of the replica index.

\section{Replica trick with real replicas}

The replica trick is used in systems with quenched randomness to overcome
averaging of logarithm. The averaged free energy is defined in the replica
trick as $\beta F_{av} = -\lim\limits_{n\to0} \left(\left\langle
Z^n\right\rangle_{av} - 1\right)/n$. To investigate stability of
thermodynamic homogeneity we replicate $\nu$ times the original spin
space. The averaged free energy of such a replicated system is $ \beta
F_{\nu} = - \left\langle \ln{Z^{\nu}} \right\rangle_{av} / \nu = -
\lim\limits_{n\to0} \left(\left\langle [Z^\nu]^n \right\rangle_{av} -
1\right)/n\nu $. If the limit $n\to0$ leads to a thermodynamically
homogeneous solution the resulting free energy~$F_\nu$ should be
independent of the parameter~$\nu$. Hence thermodynamic homogeneity is
equivalent to scale invariance of the replica representation, $n\to n\nu$.

We introduce two types of replica indices in the replica trick applied to a
$\nu$ times replicated system. Real replicas we denote $a, b =
1,\ldots,\nu$ while $\alpha,\beta = 1,\ldots,n$  we use for the
mathematical ones. After averaging over the random spin-spin interactions
in the SK model we obtain at the saddle-point
\begin{equation}\label{eq:replicated-FE}
\beta f_{\nu} = \lim\limits_{n\to0}\frac 1{n\nu}\left[ \frac{\beta^2
L_{n\nu}}{2} - \ln\text{Tr} \exp{\left\{\beta^2
\sum_{\alpha\le\beta}^{n}\sum_{a\le b}^\nu
\left(Q^{ab}\right)_{\alpha\beta}  S^a_\alpha S^b_\beta + \beta
h\sum_{\alpha}^n \sum_{a}^\nu S^a_\alpha\right\}}\right]\ .
\end{equation}
We denoted $L_{n\nu}= \sum_{\alpha\le \beta}^n \sum_{a \le b}^\nu
\left(Q^{ab}\right)^2_{\alpha\beta} + n\nu/2$. Parameters
$\left(Q^{ab}\right)_{\alpha\beta} = \langle S^a_\alpha S^b_\beta\rangle$
are determined from saddle point equations \cite{SK}.

To proceed further with thermal averaging we have to determine symmetry of
the matrix $\left(Q^{ab}\right)_{\alpha\beta}$. We use the replica
symmetric ansatz and distinguish only diagonal and off-diagonal elements.
The super-diagonal elements are fixed, $\left(Q^{aa}\right)_{\alpha\alpha}
= 1$. Since the elements in the matrix of mathematical replicas are
$\nu\times\nu$ matrices, we can choose one parameter for the off-diagonal
elements in the real-replica space even if the indices for the mathematical
replicas are equal. We denote $\left(Q^{a\neq b}\right)_{\alpha\alpha}=
q_1$. In the spirit of the replica-symmetric ansatz we choose for the
off-diagonal matrix elements in the space of mathematical replicas
$\left(Q^{ab}\right)_{\alpha\neq\beta}= q_0$. For convenience we denote
$\chi = q_1 - q_0$, $q = q_0$. It is now a straightforward task to perform
averaging over the spin configurations in Eq.~\eqref{eq:replicated-FE}
using the above parameters. We arrive at
\begin{multline}\label{eq:FE-1RSB}
f_1(q,\chi;\nu) = -\frac \beta4(1-q)^2  + \frac \beta4(\nu - 1)\chi(2 q +
\chi) + \frac\beta2 \chi- \\ -\frac 1{\beta\nu}
\int_{-\infty}^{\infty}\mathcal{D}\eta\ln\int_{-\infty}^{\infty}
\mathcal{D}\lambda  \left\{2\cosh\left[\beta\left(h +
\eta\sqrt{q} +\lambda\sqrt{\chi}\right)\right]\right\}^{\nu}\ .
\end{multline}
We used an abbreviation for the Gaussian differential $\mathcal{D}\phi
\equiv {\rm d}\phi\  e^{-\phi^2/2}/\sqrt{2\pi}$.

The obtained free energy density depends formally on the scaling parameter
$\nu$. It is clear from the dependence of $f_1$ on $\nu$ in
Eq.~\eqref{eq:FE-1RSB} that it actually depends on~$\nu$ if and only if
$\chi >0$. We obtain easily from the saddle-point equations for $q$
and~$\chi$ that this happens just below the de Almeida-Thouless instability
line. Hence, the replica-symmetric solution becomes thermodynamically
inhomogeneous (free energy does not obey the Euler condition) in the
spin-glass phase. A thermodynamically inhomogeneous solution is physically
inacceptable and has to be amended. The homogeneous free energy
should be independent of $\nu$. The dependence on this
parameter is minimized in the saddle point where the free energy $f_1$
does not depend on $\nu$ at least locally. That is, we demand
$\partial f_1/\partial
\nu = 0$. In the homogeneous state this equality becomes identity, but
when we have an inhomogeneous solution this equation just determines an
"equilibrium" scaling parameter $\nu_{eq}$.

Comparing free energy \eqref{eq:FE-1RSB} where parameters $q,\chi$ and
$\nu$ are determined from the saddle-point equations with the Parisi RSB
scheme  we find equivalence with $1$RSB solution \cite{Parisi}. The
difference between these two constructions lies in their derivation and
motivation. While Parisi arrives at 1RSB solution from the demand of
maximality of the free energy, here we derived the saddle-point equations
from the effort to restore thermodynamic homogeneity (minimize
inhomogeneity). We believe that the latter construction has a more
profound physical justification.
 
The local thermodynamic homogeneity achieved by free energy
\eqref{eq:FE-1RSB} is not enough to become a thermodynamically consistent
solution. It would be if this newly won solution were independent of a new
scaling (replicating) of the corresponding spin space. To check whether
free energy \eqref{eq:FE-1RSB} is homogeneous, i.~e., independent of
global scalings of the phase space, we write the number of real replicas
as a product $\nu=\nu_1\nu_2$, whereby e.~g. $\nu_1$ be determined from
stationarity equations resulting from Eq.~\eqref{eq:FE-1RSB}. We then have
two types of real replicas. The first one with indices $a,b=1,\ldots,\nu_1$
determines the $1$RSB solution, while the other with indices
$a',b'=1,\ldots,\nu_2$ serves to a replication of the phase space in which
free energy  $f_1$ was derived. Using again the replica-symmetric ansatz
for averaging we obtain only three parameters that we denote  $(Q^{a\neq
b,a' a'})_{\alpha\alpha} = q + \chi_1,(Q^{ab, a'\neq b'})_{\alpha\alpha} =
q + \chi_2, (Q^{ab, a'b'})_{\alpha\neq\beta} = q$. It is again a
straightforward task to derive an explicit representation for the free
energy in the $\nu_2$-times replicated spin space leading to
Eq.~\eqref{eq:FE-1RSB}. Not surprisingly we find that the resulting free
energy  formally corresponds to the Parisi $2$RSB solution.
%

We can proceed with phase-space scalings in each higher hierarchical state
with $q,\chi_1,\ldots,\chi_K$ and $\nu_1,\ldots,\nu_K$ as order
parameters. We would end up with the Parisi full (infinite) RSB scheme.
However, this is not the objective of this construction. We should produce
at the end a solution that is globally thermodynamically homogeneous,  the
limit $n\to0$ is scale independent.

\section{Global thermodynamic homogeneity}

Determining all the physical parameters $q,\chi_1,\ldots,\chi_K$ and the
geometric ones $\nu_1,\ldots,\nu_K$ from the saddle-point equations of the
$K$RSB free energy, we are left with the number of hierarchies $K$ as the
only free parameter in this construction. This number will be fixed by
reaching the global thermodynamic homogeneity. The solution becomes
globally homogeneous when the free energy does not decay into a~higher
nontrivial hierarchy, i.~e., when the number of parameters
$\chi_1,\ldots,\chi_K$ does not increase when the phase space (replica
index) is scaled.

\begin{figure}
\includegraphics[width=7cm]{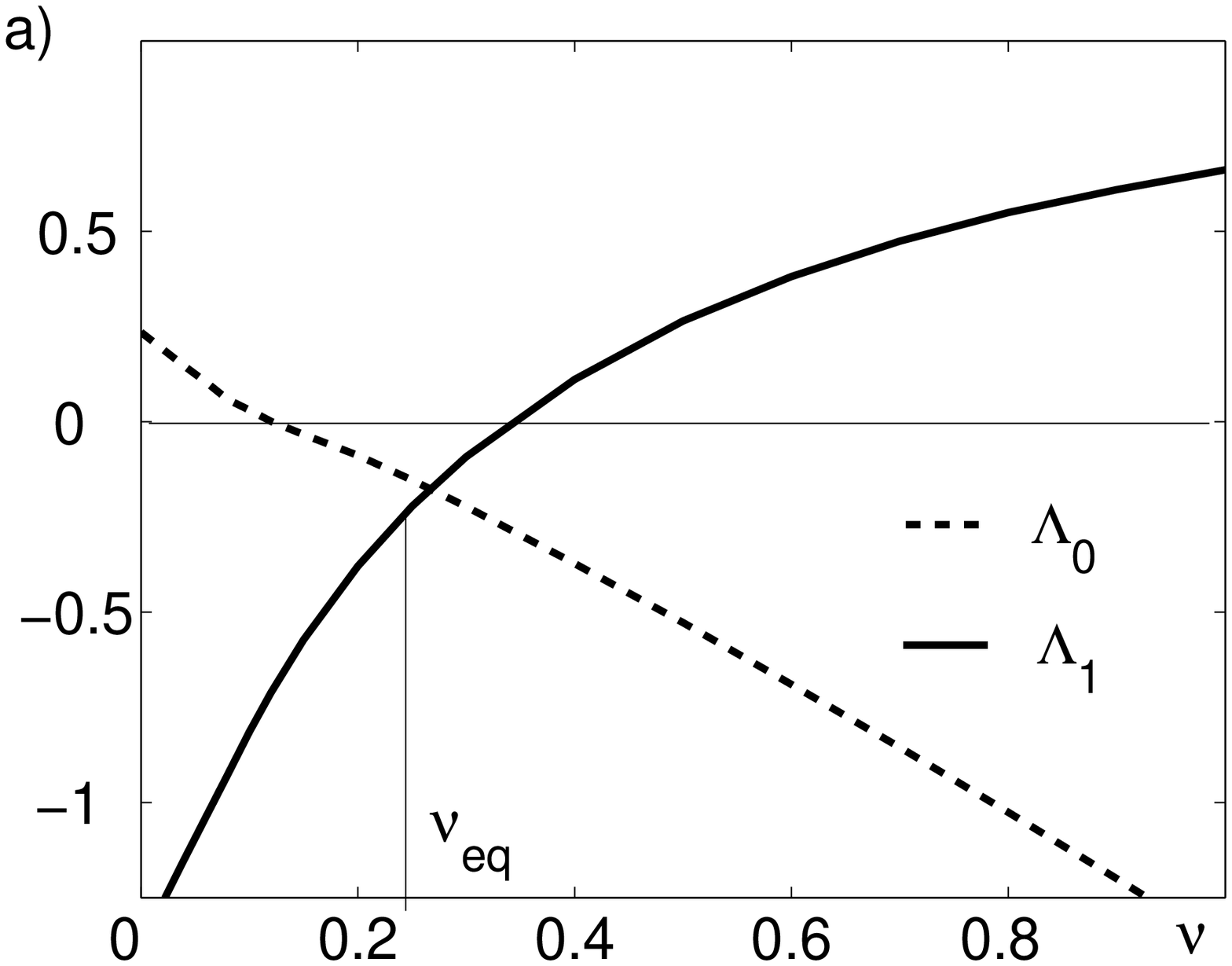}
\includegraphics[width=7cm]{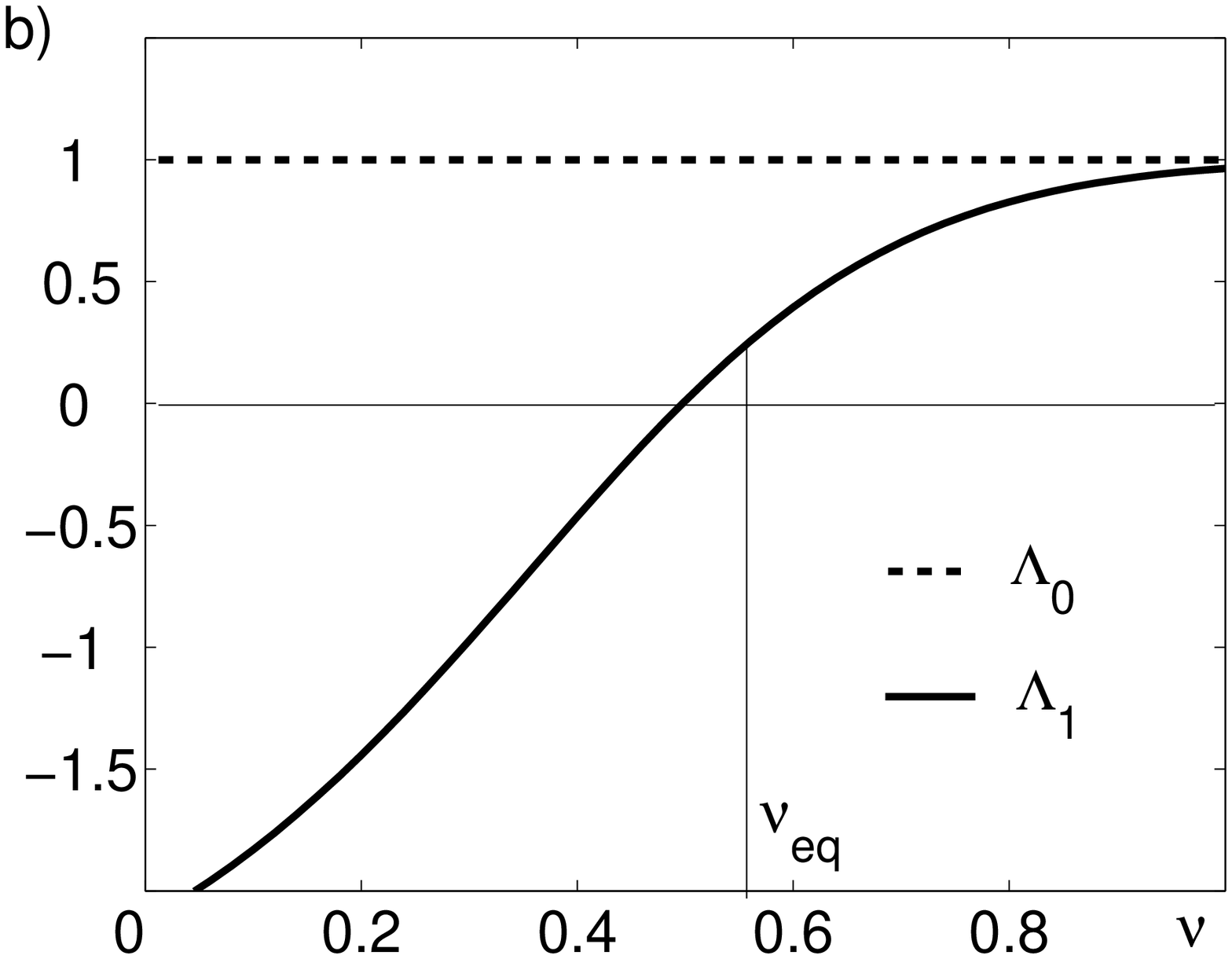}
\caption{\label{fig:1-instab}Stability parameters
$\Lambda_1$ and $\Lambda_0$ for a) SK model at $T/J=0.25$ and b)
$3$-spin model at $T/J=0.4$. Equilibrium $\nu_{eq}$ is indicated showing
that the 1RSB in the SK model in inhomogeneous, while in the $p$-spin
model is globally homogeneous.}
\end{figure}

Having the 1RSB free energy \eqref{eq:FE-1RSB} we have two possibilities
how a 2RSB solution with $\chi_1 > \chi_2 >0$ may arise. First,
$\chi\to\chi_2$ and $\chi_1=\chi_2$ becomes unstable. It happens if
\begin{subequations}\label{eq:stability-1RSB}
\begin{equation}\label{eq:stability-Delta}
\Lambda_1 =1- \beta^2\left\langle \left\langle\rho_\nu \left(1 -
t^2\right)^2 \right\rangle_\lambda\right\rangle_\eta < 0\ ,
\end{equation}
where we denoted $t\equiv \tanh\left[\beta(\eta\sqrt{q} + \lambda\sqrt{\chi})
\right]$, $\langle
X(\lambda) \rangle_{\lambda} = \int_{-\infty}^{\infty}\mathcal{D}\lambda
\ X(\lambda)$ and $\rho_\nu\equiv \cosh^\nu\left[\beta( \eta\sqrt{q} +
\lambda\sqrt{\chi}) \right]/\left\langle \cosh^\nu\left[\beta(
\eta\sqrt{q} + \lambda\sqrt{\chi}) \right]\right\rangle_\lambda$.
Notice that condition \eqref{eq:stability-Delta}
is formally equivalent to the smallest replicon for 1RSB. Second,
$\chi\to\chi_1$ and $\chi_2=0$ becomes unstable,
which is the case if
\begin{equation}\label{eq:stability-chi2}
\Lambda_0 = 1 - \beta^2\left\langle \left[ 1 - \langle\rho_\nu
t^2\rangle_\lambda +  \nu \left(\langle\rho_\nu t^2\rangle_\lambda -
\langle\rho_\nu
t\rangle_\lambda^2\right)\right]^2\right\rangle_\eta<0\ .
\end{equation}
\end{subequations}
Hence free energy \eqref{eq:FE-1RSB} becomes thermodynamically homogeneous
if both numbers $\Lambda_1$ and $\Lambda_0$ from
Eqs.~\eqref{eq:stability-1RSB} are nonnegative for the equilibrium scaling
parameter~$\nu_{eq}$. It is never the case in the SK model but it may
happen above the Gardner temperature in the $p$-spin infinite-range model
 as illustrated in Fig.~\ref{fig:1-instab}.
In the p-spin model the
stability parameters are only slightly modified
$\Lambda_1 = 1-\beta^2 p(p-1) q_1^{p-2}\langle \ldots \rangle_{\eta}/2$,
$\Lambda_0 = 1 - \beta^2 p(p-1)q_0^{p-2}\langle \ldots
\rangle_{\eta}/2$. For $p=2$ it is the SK, for $p>2$ there is always
$q_0 = 0$ leading to $\Lambda_0 = 1$ \cite{pspin}.

To conclude, we demonstrated that the demand of thermodynamic homogeneity
can be build in into the replica trick as scale invariance of the replica
index $n$. Successive scalings of the replica index lead to a formal
dependence of the solution on the scaling parameter. The replica symmetric
solution depends on the scaling parameter virtually only in the spin glass
phase and hence gets unstable. The RSB scheme of Parisi  simultaneously
with conditions for the global homogeneity (scale invariance) are then
derived from successive scalings of the replica index by minimizing the
incurred inhomogeneity (demanding local homogeneity or scale invariance).


\section*{Acknowledgements}
The work on this problem was supported in part by Grant IAA1010307 of
the Grant Agency of the Academy of Sciences of the Czech Republic and
the ESF Programme SPHINX.

%

\end{document}